\documentclass{article}
\usepackage{spconf,amsmath,graphicx}

 \usepackage{subfigure}
\title{Generative Model Based Highly Efficient Semantic Communication Approach for Image Transmission}
\name{$\text{Tianxiao Han}^{1}$, $\text{Jiancheng Tang}^{1}$, $\text{Qianqian Yang}^{1}$, $\text{Yiping Duan}^{2}$, $\text{Zhaoyang Zhang}^{1}$, $\text{Zhiguo Shi}^{1}$}
\address{$\:^{1}$College of Information Science and Electronic Engineering, Zhejiang University, Hangzhou, China\\ 
$\:^{2}$Department of Electronic Engineering, Tsinghua University, Beijing, China
}

\begin{document}
%
\maketitle
\begin{abstract}

Deep learning (DL) based semantic communication methods have been explored to transmit images efficiently in recent years. In this paper, we propose a generative model based semantic communication to further improve the efficiency of image transmission and protect private information. In particular, the transmitter extracts the interpretable latent representation from the original image by a generative model exploiting the GAN inversion method. We also employ a privacy filter and a knowledge base to erase private information and replace it with natural features in the knowledge base. The simulation results indicate that our proposed method achieves comparable quality of received images while significantly reducing communication costs compared to the existing methods. 
\end{abstract}
\section{Introduction}
\label{sec:intro}
With the explosion of visual content in our daily life, such as pictures, movies, and ever the undergoing metaverse applications, the efficient transmission of images becomes more important to improve the performance and experience of the current wireless communication systems. Semantic-oriented communication methods have brought a tremendous improvement in transmission efficiency because of the joint extraction and compression of the semantic information from the source leveraging deep learning methods. In \cite{bourtsoulatze2019deep}, the authors are the first to apply the autoencoder model for joint source and channel coding and achieve better transmission efficiency of images. The authors \cite{zhang2022wireless} proposed a multi-level Semantic communication system to extract both the high-level and low-level semantic information of an image. However, the compression ratio of all these works ranges between $1/60$ and $1/20$, which may be insufficient due to the explosion of demands in the near future. A task-specific semantic communication system was proposed in \cite{jankowski2020wireless} for image transmission targeting person re-identification, which achieves a compression ratio of $1/128$. But it considers the specific task and is not able to reconstruct the image at the receiver. In this paper, we aim to propose a highly efficient semantic communication method for image transmission with a low compression ratio while preserving the perceptual quality. 

The generative adversarial network (GAN) models have the capability to generate high dimensional contents from a low dimension vector. Some existing works on semantic communication systems exploit this property of GAN in the image transceivers to achieve a small compression ratio. The authors in \cite{huang2021deep} use conditional GAN at the receiver to reconstruct the image, which requires the transmitter to send the semantic segmentation labels and the residual images, resulting in additional transmission overhead. The authors in \cite{wang2022perceptual} use GAN along with the Deep JSCC model \cite{bourtsoulatze2019deep} to achieve better transmission efficiency and perceptual metrics. However, the smallest compression ratio achieved so far is $1/50$.

In this paper, we exploit the recent advance of interpretable generative models \cite{karras2019style} \cite{liu2020semanticgan}, in particular, semantic StyleGAN proposed in \cite{shi2022semanticstylegan}, to extract disentangled semantic information from input images and further improve the transmission efficiency. This is achieved by training the network with segmentation labels in a supervised manner. For example, the different parts in latent codes obtained by semantic StyleGAN trained on the human face dataset present different parts of a human face. We also take into account the privacy concern of the image transmission. Note that the straightforward method is to erase the private part and transmit the masked images directly. However, it results in unnatural perception to the viewer because the removal of private parts may make the generated images fall out the distribution of natural images \cite{meden2021privacy}. 
We tackle this issue by ensuring the modified latent code to be within the latent space \cite{creswell2018generative}, in order to protect private information while reconstructing the natural image simultaneously. 


The contributions of this paper can be summarized as follows: (i) A generative model based semantic communication framework is proposed with the inversion method of semantic StyleGAN to extract semantic information and a normalizing flow model with its inversion to help achieve an extremely small compression ratio, i.e., $1/3072$, which is only $1\%$ of existing works. (ii) We demonstrate that manipulating the latent codes of semantic StyleGAN with a privacy filter and a knowledge base helps reconstruct images naturally while protecting privacy. (iii) As numerical experiments show, we achieve a comparable quality of received images while significantly reducing communication costs compared to the existing methods, especially under harsh communication conditions. 

\section{SYSTEM MODEL}
\label{sec:format}

\begin{figure*}[htb]

  \centering
  \centerline{\includegraphics[width=15.5cm]{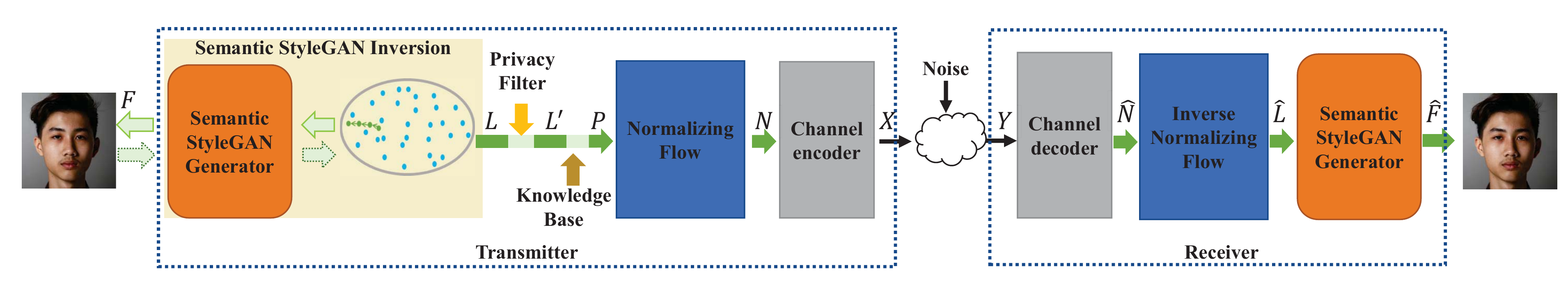}}

%
\caption{The overall architecture of the proposed generative model based semantic communication system for image transmission.}
\label{fig:model}
\end{figure*}

In this section, we present the system model of the considered semantic communication system for privacy-preserving image transmission. We also introduce the metrics to evaluate the performance of the proposed model. 


\subsection{Transmitter}
As shown in Fig. \ref{fig:model}, the input image $\boldsymbol F$ is first transformed into a latent representation $\boldsymbol L$ by Semantic StyleGAN with the inversion method. Then a privacy filter is applied to filter out the privacy information and obtain a privacy-preserving latent representation $\boldsymbol {L^{\prime}}$. Then we      
    select features in the semantic knowledge base to replace the erased information and get $\boldsymbol P$, which helps the natural reconstruction of the images with the private information replaced. Normalizing flow transformation \cite{kobyzev2020normalizing} is applied on $\boldsymbol P$ to get a latent representation $\boldsymbol N$ with a distribution that are optimized for transmission and reconstruction. Finally, the channel encoder maps the $\boldsymbol N$ into symbol sequences $\boldsymbol X$ to be transmitted  that can be robust to imperfect channel.
\subsection{Receiver}
The received signal at the receiver is given by
\begin{equation} {\boldsymbol Y=\boldsymbol h \ast \boldsymbol X+\boldsymbol n}, \label{channel}  \end{equation}
where $\boldsymbol h$ represents the channel coefficients, and $\boldsymbol n$ denotes the Gaussian noise of channel. The received signal, $\boldsymbol Y$, is then mapped back into the latent semantic representation $\widehat{\boldsymbol L}$ by a channel decoder and the inverse normalizing flow, which is then converted into the reconstructed images $\widehat{\boldsymbol F}$ by the generator of Semantic StyleGAN.


\subsection{Metrics}
We evaluate our system performance with both the objective and subjective metrics. For the objective evaluation, we compare the input images and reconstructed images at the pixel level with Peak Signal to Noise Ratio(PSNR). PSNR is calculated by the Mean Square Error(MSE) between the two images, denoted by
\begin{equation}
PSNR=10\times log_{10}\frac{Max(F,\widehat{F})^2}{MSE (F,\widehat{F})} ,
\label{PSNR}
\end{equation}
where $Max(F,\widehat{F})$ is the maximum possible pixel value of the image $ \boldsymbol F$ and $\boldsymbol{ \widehat{F}}$.
We also use the deep learning based perceptual metric, LPIPS\cite{zhang2018unreasonable}, to evaluate our system performance. LPIPS evaluate the reconstructed image from human perceptual perspective by computing the distance between the original and the reconstructed images in the feature space derived by the pretrained VGG network.  We have 
\begin{equation}
LPIPS(F,\widehat{F}) =  \sum_l{\frac{1}{H_l W_l}} \sum_{i,j} \Vert c_l \odot (f^l_{ij} - \hat{f}^l_{ij}) \Vert^2_2,
\label{eq:lpips}
\end{equation}
where $f^l$ and $\hat{f}^l$ denote the normalized latent feature map output by layer $l$ of VGG network with $F$ and $\widehat{F}$ as input, respectively. $H_l$, $W_l$ and subscript ${ij}$ denote the height, width and $(i,j)$th element of the feature map, respectively. Notation $\odot$ denotes the scale operation and $c_l$ represents the pretrained weights for the features in layer $l$ which is used to scale the feature map.



\section{Proposed method}
In this section, we present the details of the proposed model depicted in Fig. \ref{fig:model}. We first introduce the proposed generative models based transmission scheme and then the privacy-preserving mechanism to protect the privacy of the transmitted images by utilizing a privacy filter and a knowledge base.


\subsection{Generative Model Based Transmission}
 
In this paper, we adopt semantic StyleGAN\cite{shi2022semanticstylegan}, which generates a latent representation of the input images with different parts representing different semantic information. This is achieved by training this generative model with semantic segmentation labels as supervised signals, where a dual-branch discriminator models the joint distribution of RGB images and semantic segmentation labels. Due to this property, we exploit the inversion method of a pretrained semantic StyleGAN \cite{abdal2019image2stylegan} to extract the disentangled semantic information in latent spaces from the input image. The inversion method works as follows: it starts with an initial vector in the latent space, which is then input into the Semantic StyleGAN to generate an image. Then we calculate the MSE loss between the generated image and input image $\boldsymbol F$, by which we derive gradient descents of the vector in the latent space. After several iterations, we the optimal latent code of the input image in the latent space and output this code as the latent representation $\boldsymbol L$.

However, it is hard to combine the semantic StyleGAN with the channel encoder and channel decoder in an end-to-end method during training, because the training process of semantic StleGAN, which consists of 28 feature generators with 10 FC layers each, along with a stack of convolution layers, is quite slow to train and often fails to converge. Therefore, we need diffeomorphic mapping \cite{milnor1997topology} from the semantic latent code to a new and easily trainable representation. We adopt a normalizing flow model \cite{kobyzev2020normalizing} to learn an invertible mapping between latent spaces of semantic StyleGAN and a group of Gaussian distributions. At the receiver side, we then use the inverse normalizing flow model to map the received $X$ back to the semantic latent representation $\boldsymbol{\widehat{L}}$, by taking advantage of the invertible property of normalizing flow in computational efficiency. More specially, we use the RealNVP model proposed in \cite{dinh2016density}, which consists of 8 coupling layers, and each coupling layer consists of 6 FC layers. The RealNVP is quite simple but efficient to train compared to semantic StyleGAN. As revealed by experiments, the usage of normalizing flow helps improve the quality of reconstructed images.

We then employ relatively simple channel encoder and decoder, each consisting of two cascaded FC layers, are used to map latent codes into symbol sequences $\boldsymbol X$ at the transmitter, and map received symbol sequences $\boldsymbol Y$ back to latent representation $\widehat{\boldsymbol N}$ at the receiver.

\begin{figure}[htb]

\begin{minipage}[b]{.32\linewidth}
  \centering
  \centerline{\includegraphics[width=2cm]{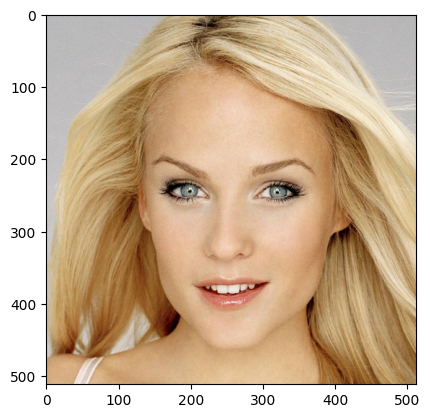}}
  \centerline{(a) Input image from } \medskip
  \centerline{Celeb-HQ dataset \cite{CelebAMask-HQ}} \medskip
\end{minipage}
\hfill
\begin{minipage}[b]{0.32\linewidth}
  \centering
  \centerline{\includegraphics[width=2cm]{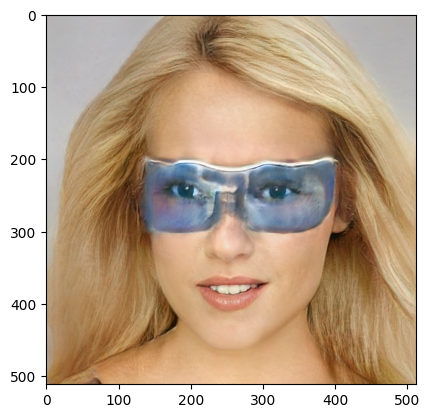}}
  \centerline{(b) Reconstructed}\medskip
  \centerline{image w/o KB} \medskip
\end{minipage}
\hfill
\begin{minipage}[b]{0.32\linewidth}
  \centering
  \centerline{\includegraphics[width=2cm]{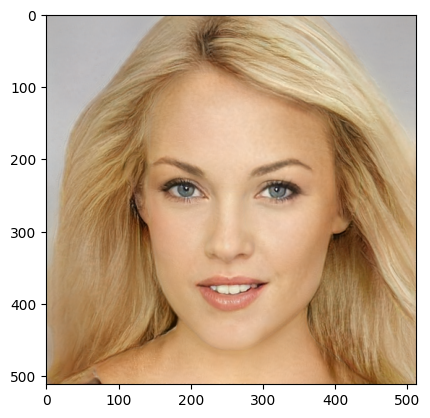}}
  \centerline{(c) Reconstructed }\medskip
  \centerline{image with KB} \medskip
\end{minipage}
\caption{Generated images with and without the knowledge base(KB).}
\label{fig:privacy}
\end{figure}
\subsection{Privacy preserving mechanism}



We include a privacy preserving mechanism by exploiting a privacy filter and a knowledge base. Since private parts and non-privacy parts in latent representation $L$ for are disentangled, we design a privacy filter that adds a bias to the private parts and obtain a privacy-preserving representation $\boldsymbol {L^{\prime}}$. The bias in the privacy filter is set to ensure the Euclidean distance between the reconstructed image and the input image to be larger than a given threshold, i.e.,
\begin{equation}
d(\boldsymbol {L^{\prime}},\boldsymbol L) >  \lambda_1 \ast \boldsymbol L, 
\end{equation}
where $\lambda_1$ is a task-specific parameter. 

However, the randomly added bias may cause the latent representation $\boldsymbol {L^{\prime}}$ to fall into an out-of-distribution area and generate an unreasonable feature, as the Fig. \ref{fig:privacy} shows. Therefore, we further exploit a semantic knowledge base (KB) to replace the private parts with certain latent codes that generate a natural image suitable for human perception. The semantic KB uses all the real images in the dataset to obtain the average latent codes $\boldsymbol L_{m}$, which is a guideline for a reasonable generation. Then we add another bias to $\boldsymbol {L^{\prime}}$ and obtain a constrained representation $\boldsymbol P$. The added bias is chosen such that the distance between the reconstructed image and the natural images while forming an upper bound for the distance between the reconstructed image and the privacy-protected images, i.e.,



\begin{equation} 
d(\boldsymbol P,\boldsymbol {L^{\prime}}) < \lambda_2 \ast d(\boldsymbol P,\boldsymbol L_{m}), \label{upper} 
\end{equation} where $\lambda_2$ is less than $1$ and can control the diversity of reconstructed image. It is noted that the bias added start with a small value and gradually increase until \eqref{upper} is satisfied. 


\subsection{Training Methods}

We use a two-stage training method, which achieves better performance. In the first stage, we train the normalizing flow and ignore the channel encoder and decoder by directly inputting the output of the normalizing flow at the transmitter to the inverse normalizing flow at the receiver. We use MSE loss functions between the reconstructed image and the input image, together with the normalizing flow loss. In the second stage, We train the whole network together with an additional MSE loss function between $\boldsymbol N$ and the received $\widehat{ \boldsymbol{N}}$ and use a training method during which the noise gradually increases to prevent the model from crashing or collapse at the beginning.

\section{Simulation Results and Analysis}
In this section, we evaluate the performance of the proposed semantic communication system for image transmission in terms of the transmission efficiency and privacy preserving. For simplicity, We assume the AWGN channel. We use the Celeb-HQ dataset \cite{CelebAMask-HQ} for training and testing, which is a human face dataset with semantic labels. We use the existing semantic communication approach by \cite{bourtsoulatze2019deep}, referred to as Deep JSCC, as the benchmark approach to compare to. We note that all approaches are trained and tested with the same datasets. During training, we set channel conditions with SNR varying randomly from 5dB to 10dB. The compression ratio is also defined in \cite{bourtsoulatze2019deep}, which is the ratio of the dimension of vectors ($k$) to be transmitted and the size 
 of the input images ($n$).  

\begin{figure}
\includegraphics[width=0.3\textwidth]{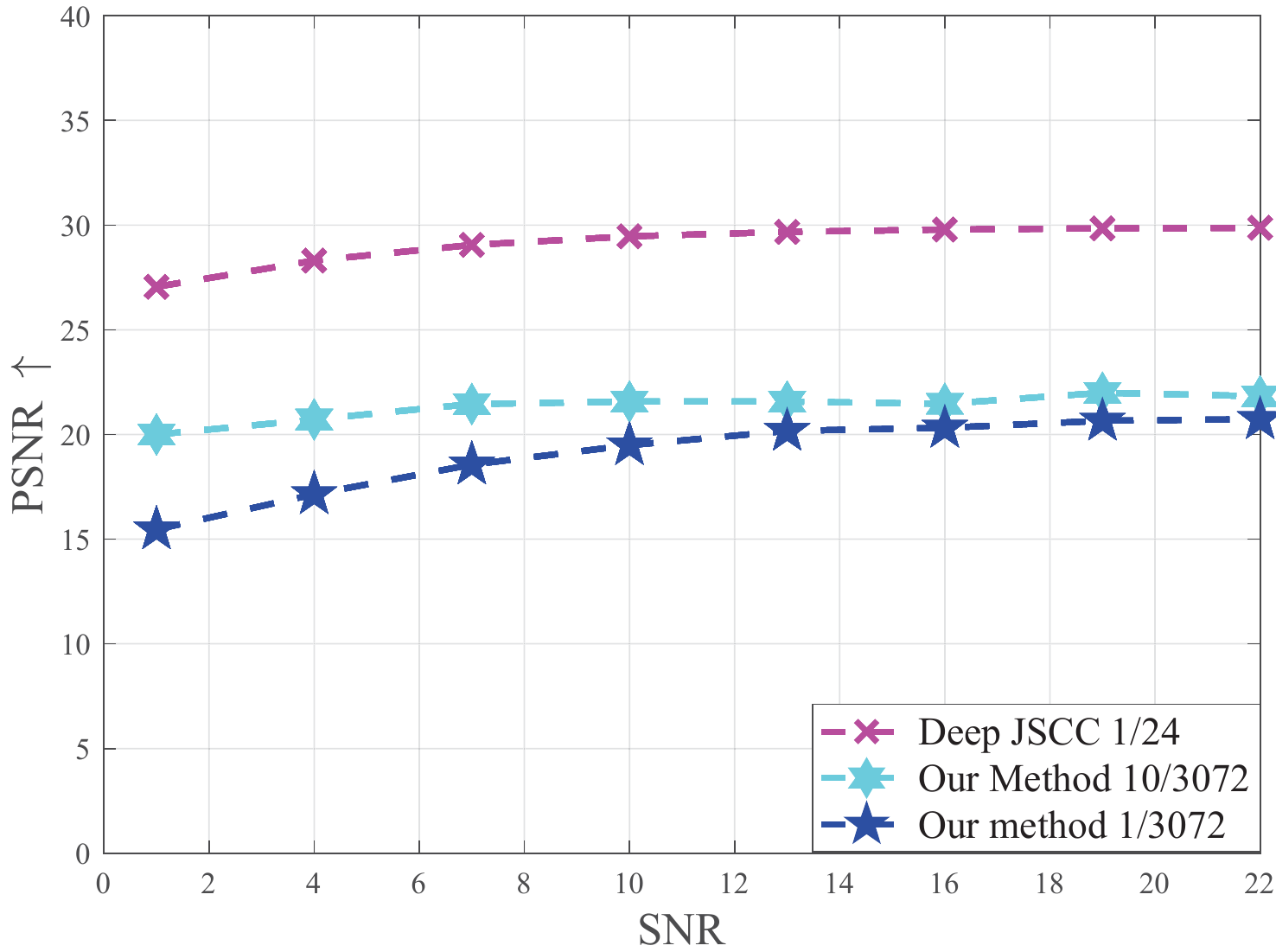}
\centering 
\caption{PSNR versus SNR for different approaches.Our methods' compression ratio are $1/3072$ and $10/3072$, while the compression ratio of Deep JSCC is $1/24$.}  
\label{PSNR1}  
\end{figure}

\begin{figure}
\includegraphics[width=0.3\textwidth]{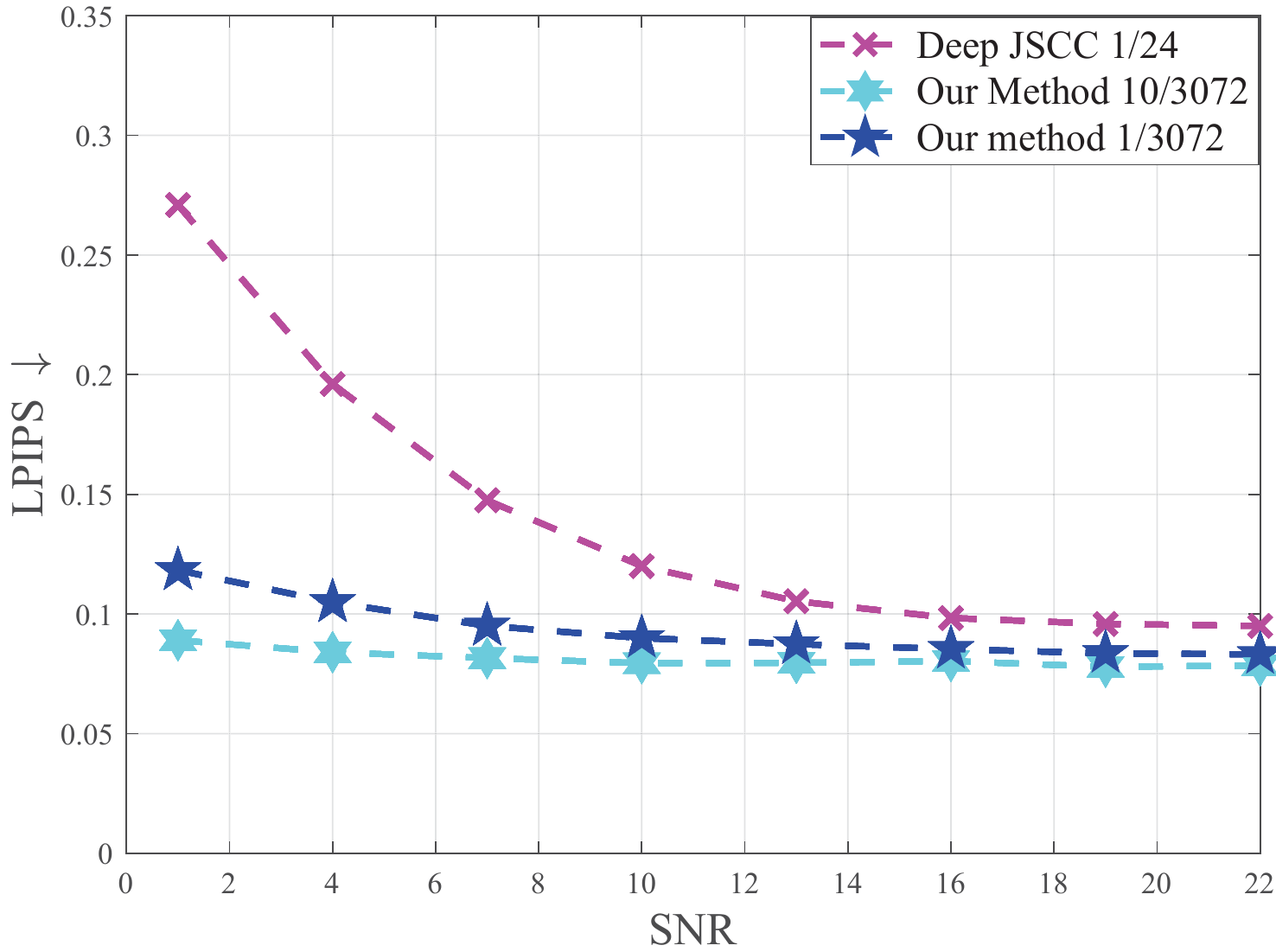}
\centering 
\caption{LPIPS versus SNR for different approaches. Our methods' compression ratio are $1/3072$ and $10/3072$, while the compression ratio of Deep JSCC is $1/24$}  
\label{LPIPS}  
\end{figure}

\begin{figure}
\includegraphics[width=0.3\textwidth]{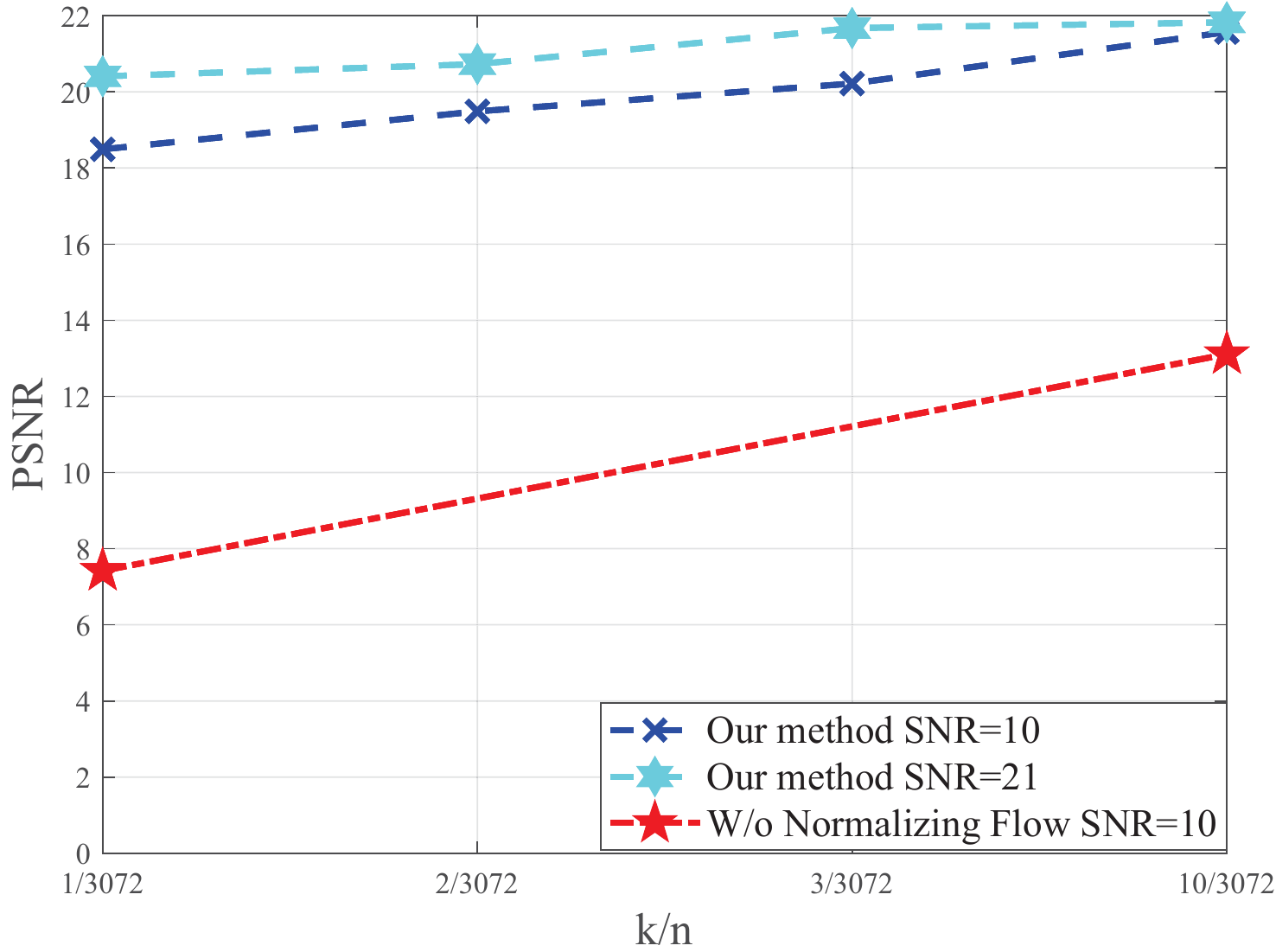}
\centering 
\caption{PSNR versus k/n and the PSNR without normalizing flow.}  
\label{kn}  
\end{figure}

The performance comparison of different approaches in terms of PSNR and LPIPS is presented in Fig. \ref{PSNR1} and Fig. \ref{LPIPS}. Higher PSNR presents better pixel level restoration, while smaller LPIPS scores represent better objective perception restoration. We can observe that with only $1\%$ the compression ratio of the existing method, our method achieves better LPIPS scores, and slightly worse PSNR. This demonstrates that semantic information we extract and transmit is highly compact and preserve the perceptual contents. We also observe that PSNR and LPIPS by our approach stay almost the same under different SNRs, which illustate the robustness of the proposed semantic communication system to noisy channel. We also compare the numerical results achieved by the proposed method with and without the normalizing flow in Fig. \ref{kn}, which proves the benefits of normalizing flow in improving the quality of image transmission.

\begin{figure}{}
\centering
\subfigure[]{\includegraphics[width=1.5cm]{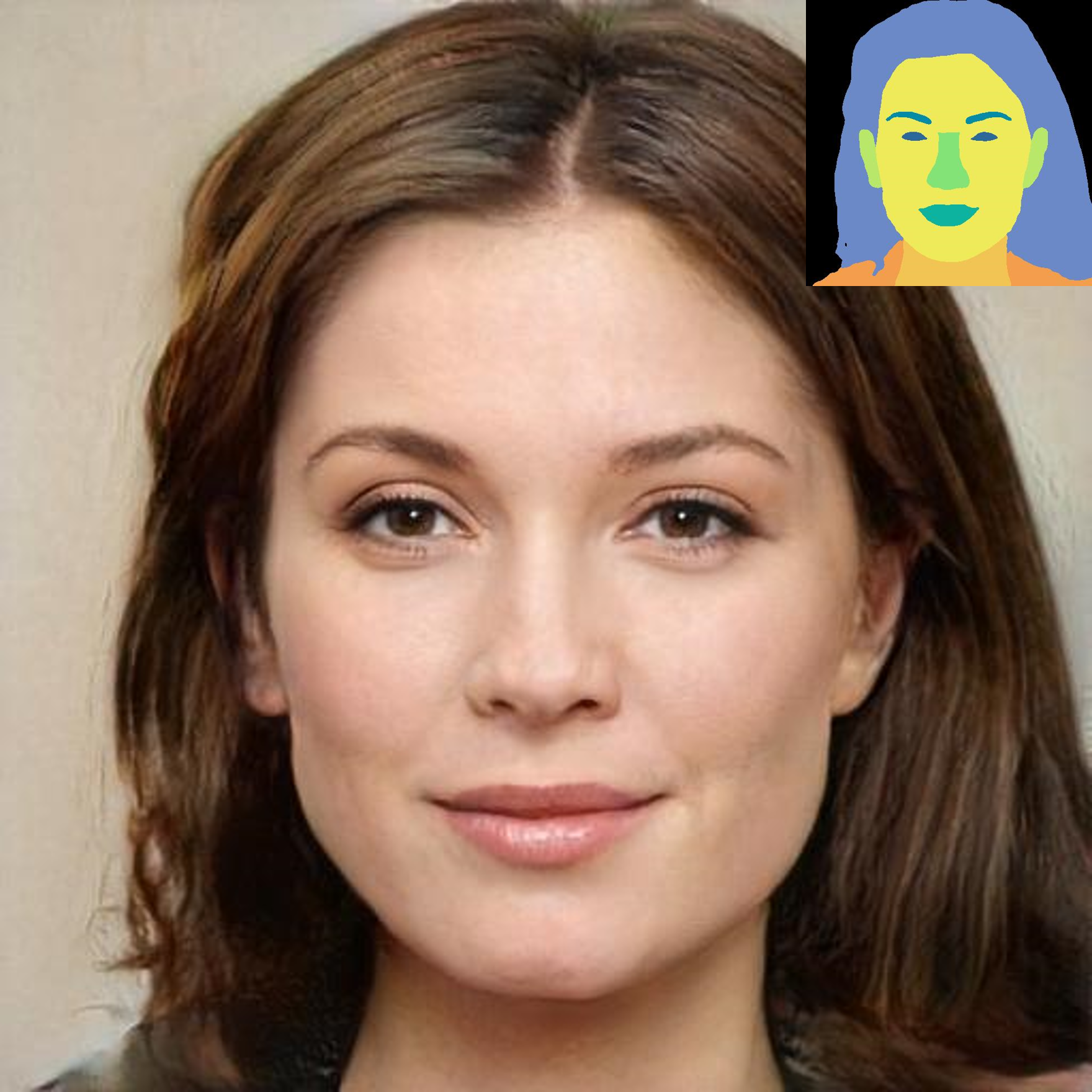}} 
\subfigure[]{\includegraphics[width=1.5cm]{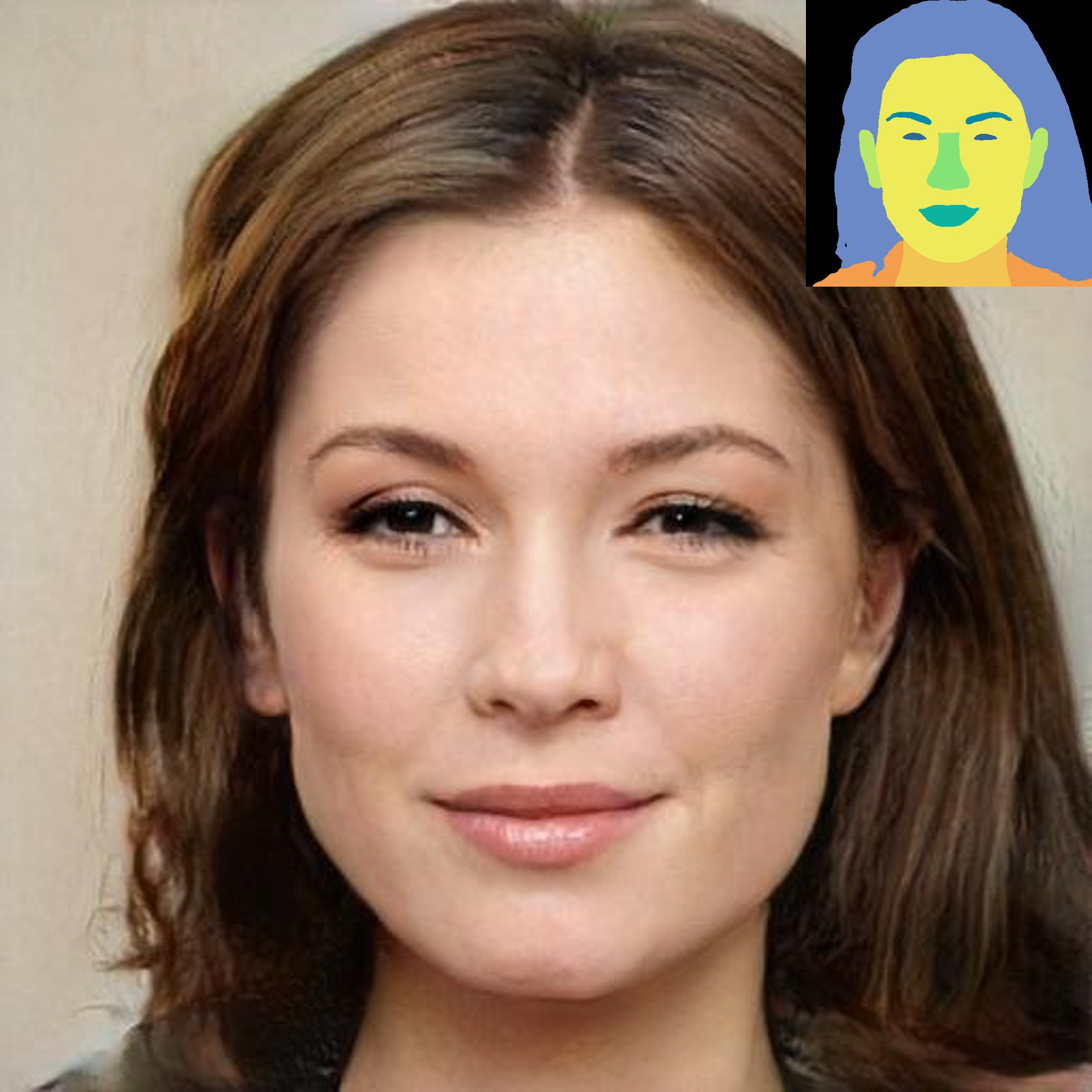}}
\subfigure[]{\includegraphics[width=1.5cm]{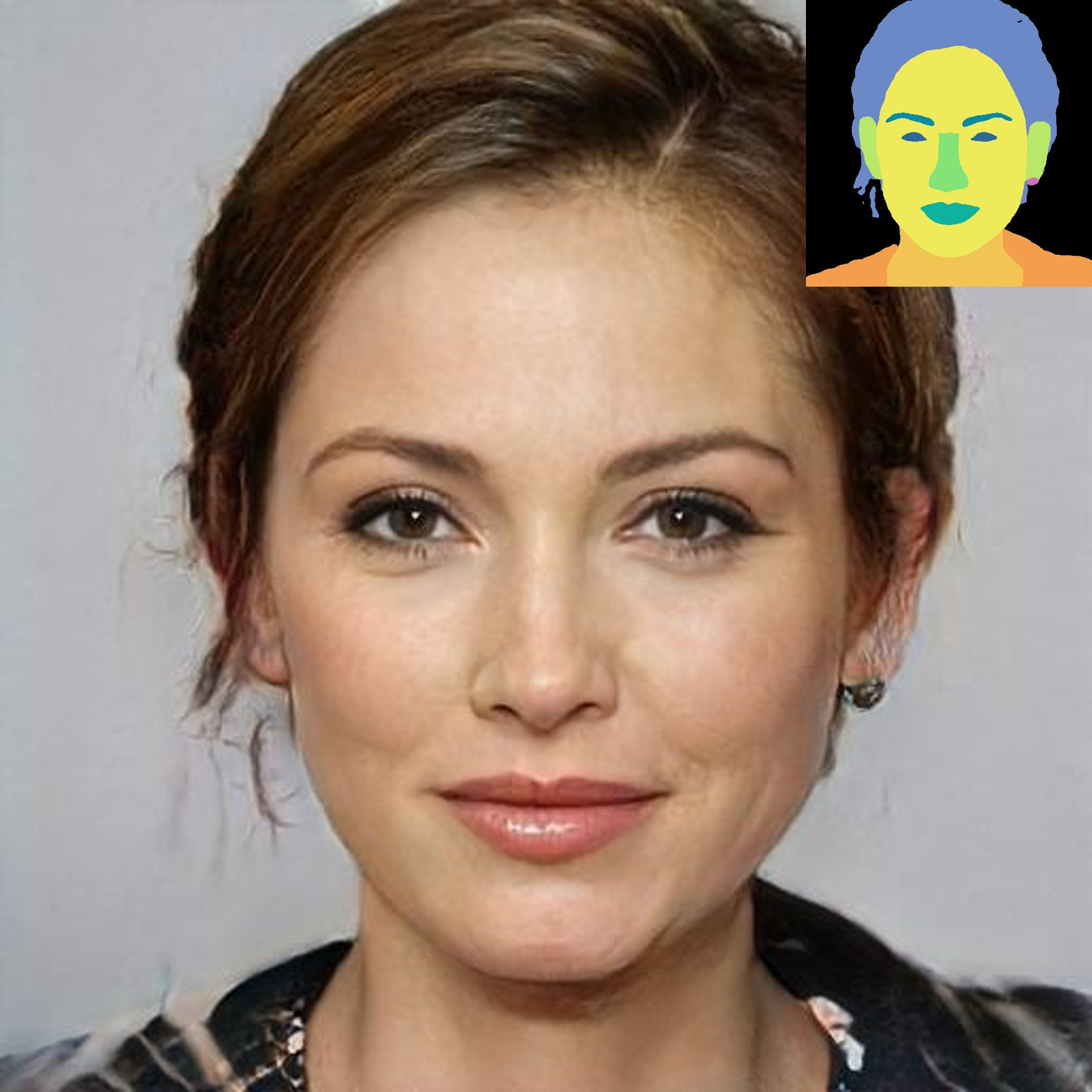}}
\subfigure[]{\includegraphics[width=1.5cm]{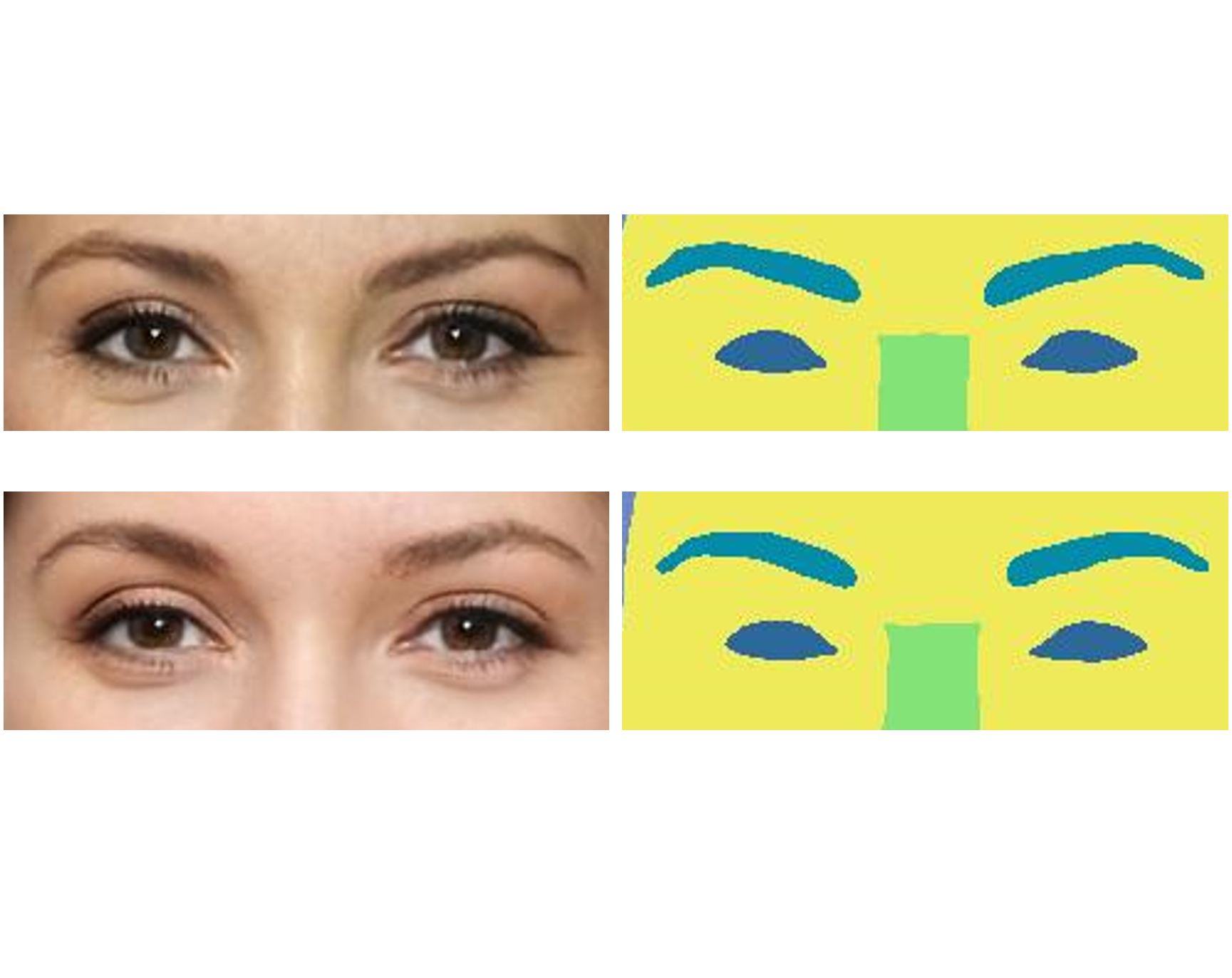}}

\caption{Example of the privacy protection. } 
\label{fig:demo}
\end{figure}

We also evaluate the privacy filter's effect in Fig. \ref{fig:demo}, where Fig. \ref{fig:demo}(a) is the original image. By assuming the eyes are the private information to protect, and we can get Fig. \ref{fig:demo}(b), where all the other parts stay the same while the eyes are different. For another case, assuming all the features are private information except eyes, and we can get Fig. \ref{fig:demo}(c). By comparing the eyes of the input image and reconstructed image, as shown in Fig. \ref{fig:demo}(d), we can see that similar eyes are kept while all the other information is changed. This shows the disentanglement of different semantic parts helps the protection of privacy.

\section{Conclusions}



In this paper, we proposed a novel generative model based image transmission approach with the ability to significantly reduce the transmission overhead while preserving the private information. We utilized the inversion method of Semantic StyleGAN to acquire the disentangled latent codes of input images, and proposed a privacy filter that modifies the latent codes with the Euclidean distance rule with the help of the knowledge base. Simulation results demonstrated that our proposed method achieves significantly better transmission efficiency with less than $1\%$ of the compression ratio by the existing method while achieving the comparable reconstruction quality and able to preserve private information.

\vfill\pagebreak

\bibliographystyle{IEEEbib}
\bibliography{strings,refs}

\end{document}